\def\draftversion{false}

\RequirePackage{ifthen}
\ifthenelse{\equal{\draftversion}{true}}{
  \documentclass[aps,prb,galley,epsfig,amsmath,showpacs]{revtex4-1}
}
{
  \documentclass[aps,prb,twocolumn,epsfig,amsmath,showpacs]{revtex4-1}
}

\ifthenelse{\equal{\draftversion}{true}}{
  \marginparwidth 2.7in
  \marginparsep 0.5in
  \newcounter{comm} 
  \def\commnext{\stepcounter{comm}}
  \def\commtext{{\bf\color{blue}[\arabic{comm}]}}
  \def\commmar{{\bf\color{blue}[\arabic{comm}]}}
  \def\hym#1{\commnext\marginpar{\small HY\commmar: #1}\commtext}
  \def\scm#1{\commnext\marginpar{\small SC\commmar: #1}\commtext}
}{
  \def\hym#1{}
  \def\scm#1{}
}

\usepackage{epsfig}
\usepackage{dcolumn}
\usepackage{bm}
\usepackage{color}
\usepackage{multirow}

\begin{document}
\title{Calculation of the specific heat of optimally K-doped BaFe$_2$As$_2$}

\author{Hyungju Oh}
\email[Email:\ ]{xtom97@civet.berkeley.edu}
\author{Sinisa Coh}
\author{Marvin L. Cohen}
\affiliation{Department of Physics, University of California at Berkeley 
and Materials Sciences Division, Lawrence Berkeley National Laboratory, 
Berkeley, California 94720, USA}

\date{\today}

\begin{abstract}
The calculated specific heat of optimally K-doped BaFe$_2$As$_2$ in
density functional theory is about five times smaller than that found
in the experiment. We report that by adjusting the potential on the iron
atom to be slightly more repulsive for electrons improves the calculated heat
capacity as well as the structural, magnetic, and electronic properties
of Ba$_{0.6}$K$_{0.4}$Fe$_2$As$_2$. Applying the same correction to
the antiferromagnetic state, we find that the electron-phonon coupling
is strongly enhanced.
\end{abstract}



\maketitle

\section{Introduction}

The discovery\cite{kamihara1} of superconductivity in LaFeAsO$_{1-x}$F$_x$ 
with a transition temperature of 26~K in 2008 
triggered unprecedented interest and further research in iron-based
superconductors.  So far, superconductivity was found in four main
families of iron-based compounds: 1111, 122, 111, and
11.\cite{stewart,oh} These iron-based materials have two phases in the
normal state: one is a paramagnetic metal and the other is an
antiferromagnetic metal.  Superconductivity emerges in both the
paramagnetic and antiferromagnetic metal phases via application of
hydrostatic pressure or carrier doping of the parent materials.
Hence, it is expected that understanding the electronic and magnetic structures 
of the metallic normal states of these systems is a needed ingredient for
unraveling the origin of the superconductivity of iron-based
materials.

Many experimental and theoretical studies have been done on the normal
states of iron-based superconductors, and a
consensus\cite{andersen,dai} has been reached in these systems that
the Coulomb interaction among the electrons is not strong enough to
induce a Mott insulating phase.  However, the Coulomb interaction plays 
an important role in determining the electric and magnetic properties.  
In the early stages of this research, theoretical insight into the properties of
these materials was gained by calculations based on density functional
theory (DFT) within the local density approximation (LDA) or
generalized gradient approximation (GGA).  However, LDA and GGA have
some limitations in describing the normal states of iron-based
superconductors.  In a paramagnetic phase, the measured mass of
low-energy quasiparticles is 2-3 times larger than that calculated
within LDA or GGA.  In addition, the measured magnitude of the ordered
moment in an antiferromagnetic phase is 2-3 times smaller than that
obtained with LDA or GGA.  Furthermore, LDA and GGA studies related to
the specific heat of these materials are not consistent with the
experimental data.  The theoretical Sommerfeld
coefficient\cite{singh,shein,hashimoto} of optimally K-doped
BaFe$_2$As$_2$ is about five times smaller than that found in the
experimental data\cite{popovich,kim,storey,kant,rotundu}.

There have been many attempts to describe electronic correlations in
these materials by combining LDA or GGA calculations with a dynamical
mean-field theory (DMFT), quasiparticle self-consistent GW (QSGW), or
the Gutzwiller method.~\cite{yin,tomczak,hansmann,wang} Using these
methods, many of the electric and magnetic properties of correlated
iron-based superconductors can be reproduced.  For example, effective
masses and Fermi surfaces (FSs) across all families of iron compounds
are well described in the framework of DFT+DMFT\cite{yin} and
QSGW\cite{tomczak}, as well as ordered moments and the fluctuations of
local moments within DFT+DMFT.~\cite{yin,hansmann} However,
a calculation of the electron-phonon coupling coefficient (which is needed
for heat capacity and superconductivity estimates) within these
approaches is non-trivial. Therefore, we use a simpler method to
calculate electronic and magnetic properties of these materials.

In this work we study the heat capacity of
Ba$_{0.6}$K$_{0.4}$Fe$_2$As$_2$ superconductor (T$_{\rm
c}$=38~K)\cite{rotter} within a semi-empirically modified GGA
potential, following studies\cite{coh} of an FeSe
monolayer. We show that one can choose a small repulsive potential
located on the iron atoms (+A term) so that the calculated specific
heat coefficient is increased from
$\gamma_n$=12~mJ~mol$^{-1}$~K$^{-2}$ in GGA to
$\gamma_n$=38~mJ~mol$^{-1}$~K$^{-2}$ in GGA+A, much closer to recent
experimental findings
($\gamma_n$=40--50~mJ~mol$^{-1}$~K$^{-2}$).\cite{popovich,kant,storey,rotundu}
The increase in $\gamma_n$ relative to GGA comes mostly from the
increased density of states (DOS) at the Fermi level and to a smaller
extent from an enhanced electron-phonon coupling.  Since
Ba$_{0.6}$K$_{0.4}$Fe$_2$As$_2$ is near a magnetic phase transition,
we also computed the heat capacity in the striped antiferromagnetic
ground state, present in the parent compound.  Just as in the
nonmagnetic calculation, we again find an increased $\gamma_n$ (from 6 to
12~mJ~mol$^{-1}$~K$^{-2}$) when +A term is added. However, unlike in
the nonmagnetic calculation, the increased $\gamma_n$ originates mostly
from increase in the electron-phonon coefficient $\lambda$.

\section{Methods}

Our calculations are based on {\it ab-initio} norm-conserving
pseudopotentials and the Perdew-Burke-Ernzerhof\cite{perdew}
functional as implemented in the SIESTA code.\cite{sanchez} Electronic
wavefunctions are expanded with pseudoatomic orbitals (double-$\zeta$
polarization). We treat the potassium doping within the virtual crystal
approximation.

Following Ref.~\onlinecite{coh} we modify the GGA
potential $V_{\rm GGA}({\bf r})$ by adding a repulsive potential on
each iron atom in the calculation,
\begin{equation}
\label{eq:correction}
V_{\rm GGA}({\bf r}) + A \sum_{i}{f ( |{\bf r}-{\bf r}_{i}| )}.
\end{equation}
Here $f(r)$ is a positive dimensionless function peaked at the nucleus
of the iron atoms (${\bf r}_i$) and the extent of $f(r)$ is comparable
with the size of $d$ orbitals in the iron atoms.  We discuss the choice of
$A$ and $f(r)$ in subsection~\ref{sec:choice}.

The GGA+A approach can be understood as a variant of the constrained
DFT (CDFT) formalism.\cite{dederichs} The CDFT approach adds a general
constraint to the density,
\begin{equation}
\label{eq_2}
\sum_{\sigma}{\int w^{\sigma}_{c}({\bf r}) \rho^{\sigma}({\bf
    r})~d{\bf r}} = N_c ,
\end{equation}
where $w_{c}({\bf r})$ acts as a weight function that defines the
constrained property.  The Kohn-Sham total energy is minimized under
the constraint from Eq.~\ref{eq_2}, by making the following functional
stationary,
\begin{equation}
W[\rho,V_c] = E[\rho] + V_c \left (\sum_{\sigma}{\int
  w^{\sigma}_{c}({\bf r}) \rho^{\sigma}({\bf r})d{\bf r}} - N_c
\right).
\end{equation}
Here $V_c$ is a Lagrange multiplier corresponding to the constraint.
Therefore, in the effective Hamiltonian of the CDFT formalism, there
is an additional potential $V_c w^{\sigma}_{c}({\bf r})$ coming from
the constraint.  Since the GGA+A potential (Eq.~\ref{eq:correction})
has the same form as the constraint potential in the CDFT approach,
GGA+A method has the same effect as constraining the number of
electrons around the iron atom.

\subsection{Choice of $A f(r)$ term}
\label{sec:choice}

Now we discuss our choice of the correction term $A f(r)$ 
appearing in Eq.~\ref{eq:correction}. 

Following previous work on the FeSe monolayer\cite{coh} we first
choose $f(r)=e^{-1.0 r^2}$ in atomic units (Bohr radius) with the
extend comparable with the size of the iron atom d-orbital. Second, we
tune $A$ from $0$ up to $A_{\rm c}$ until one of the properties of
Ba$_{0.6}$K$_{0.4}$Fe$_2$As$_2$ agrees better with the experimental
data.  We choose to tune the occupied bandwidth of the $M$-point
electron pocket since it is severely overestimated in GGA (it is
$130$~meV in GGA while $\sim$~0--10~meV in the
experiment\cite{evtushinsky,neupane}). We find that using $A_{\rm c} =
1.3$~Ry has the desired effect of tuning the M-point bandwidth
to about 4~meV.

Just as in Ref.~\onlinecite{coh} we find that the choice of $f(r)$ is
not very important for physical properties as long as it is localized
on the iron atom and $A$ is tuned for each choice of $f(r)$. For
example, using $A f(r)= 2.2 e^{-1.7 r^2}$ or $A f(r)=5.5 e^{-3.5 r^2}$
results in nearly indistinguishable band structure of
Ba$_{0.6}$K$_{0.4}$Fe$_2$As$_2$.

Using $A=A_{\rm c}$ improves not only the occupied bandwidth of the
$M$-point electron pocket but other properties of
Ba$_{0.6}$K$_{0.4}$Fe$_2$As$_2$ as well. For example, structural
parameters relevant for superconductivity (arsenic height and
iron-arsenic-iron angle)\cite{zhao,lee,kuroki,mizuguchi} are both
moved in the direction towards experimental value. Going from GGA to
GGA+A the arsenic height is increased from 1.30~\AA\ to 1.44~\AA\
while the iron-arsenic-iron angle is decreased from 112.5$^{\circ}$
to 105.9$^{\circ}$. In addition, antiferromagnetic ground state is
suppressed in GGA+A. See Table~\ref{tab:comparison} for more details.
We confirmed that the tendency for improved structural and magnetic
properties is independent of the choice of $f(r)$.

\begin{table}
\caption{\label{tab:comparison} A comparison of the arsenic height,
 iron-arsenic-iron angle, magnetic moment ($\mu_{\rm Fe}$) on iron
 atom, and the energy difference ($\Delta E$) per one iron atom
 between antiferromagnetic stripe and nonmagnetic ground state in GGA,
 GGA+A, and from experiment (Ref.~\onlinecite{rotter}) in
 Ba$_{0.6}$K$_{0.4}$Fe$_2$As$_2$.}
\begin{ruledtabular} 
\begin{tabular}{lcccc} 
 & As height & Fe--As--Fe & $\mu_{\rm Fe}$ & $\Delta E$ \\ 
 & (\AA) & ($^{\circ}$) & ($\mu_{\rm{B}}$) & (eV) \\
 \hline
 GGA+A      & 1.44 & 105.9 & 2.26 & -0.19 \\
 GGA        & 1.30 & 112.5 & 2.91 & -0.33 \\
 Experiment & 1.37 & 109.9 &     &     \\
\end{tabular}
\end{ruledtabular}
\end{table}

Finally, using $A=A_{\rm c}$ the calculated heat capacity of
Ba$_{0.6}$K$_{0.4}$Fe$_2$As$_2$ is more than three times larger as
compared to the GGA value, and in good agreement with the experimental
value. We discuss heat capacity in more detail in Sec.~\ref{sec:heat}.

\section{Electronic structure}

Now we discuss the electronic structure of
Ba$_{0.6}$K$_{0.4}$Fe$_2$As$_2$ in GGA and GGA+A.  In all of our
calculations we perform a full structural relaxation for both forces
on atoms and stresses on the cell.  We sample the Brillouin zone on a
uniform 32$\times$32$\times$32 k-point mesh.

\begin{figure} 
\epsfig{file=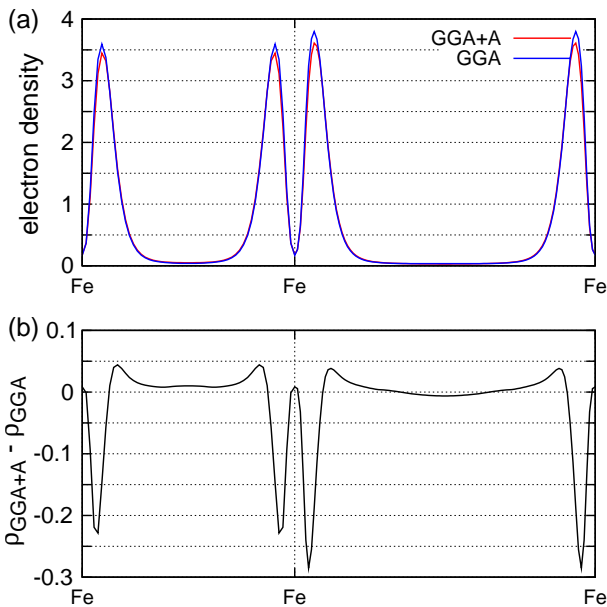,width=8cm,clip=}
\caption{The electron density of Ba$_{0.6}$K$_{0.4}$Fe$_2$As$_2$  
in the nonmagnetic state on a line between the nearest-neighbor (left)
and next-nearest-neighbor (right) iron atoms within GGA (blue) and
GGA+A (red).  Difference between GGA and GGA+A is shown in panel
(b). Densities of both semi-core (3s, 3p) and valence (3d, 4s) states
are included in our pseudopotential calculation.
\label{fig:density}
}
\end{figure}

Figure~\ref{fig:density} compares the electron density in
Ba$_{0.6}$K$_{0.4}$Fe$_2$As$_2$ in GGA and GGA+A.  From panel~b of the
figure it is clear that including the +A term transfers some of the
electronic density from the iron atom to the outer region.  The
maximal change in the electronic density is about 7~\% and it occurs
on a charge density peak near the iron atom.

\begin{figure} 
\epsfig{file=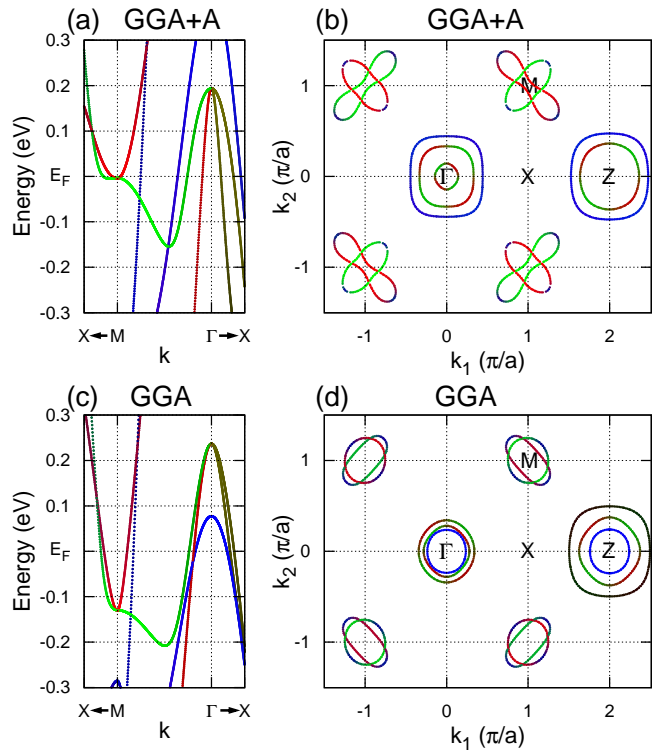,width=8.5cm,clip=}
\caption{Electronic band structures and Fermi surface of 
Ba$_{0.6}$K$_{0.4}$Fe$_2$As$_2$ in the nonmagnetic states both in
GGA+A (panels a and b) and in GGA (c,d).  Dominant orbital characters
(defined in the single-iron unit cell) are represented in blue
($d_{xy}$), red ($d_{yz}$), green ($d_{zx}$), black ($d_{z^2}$), and
yellow (both $d_{yz}$ and $d_{zx}$) color.  High symmetry points in
the Brillouin are defined in the two-iron unit cell.  Reciprocal space
axes $k_1$ and $k_2$ are perpendicular to the tetragonal $c$-axis.
\label{fig:band}
}
\end{figure}

Figure~\ref{fig:band} compares the band structure and the Fermi
surface in GGA and GGA+A. We compare these results to the experiment
in Sec.~\ref{sec:arpes}.

In the GGA case, as in a previous calculation,\cite{singh} there are
three hole pockets at the zone center ($\Gamma$), and two electron
pockets at the zone corner (M).  However, the band structures and the
Fermi surfaces in GGA+A are both quantitatively and qualitatively
different in several respects. First, the occupied bandwidth of the
$M$-point $d_{yz}$ and $d_{zx}$ electron pockets in GGA+A is 4~meV
[Fig.~\ref{fig:band}(a)], while it is 130~meV in GGA.  In addition,
the effective mass of these pockets is increased by a factor of 3--4
in GGA+A and the shape of the Fermi pocket in GGA+A is more elongated
towards the $\Gamma$ and $Z$ points.

Second, the area of the hole pockets at $\Gamma$ and $Z$ is changed in
GGA+A. Specifically, in GGA+A the size of the $d_{xy}$ hole pockets at
$\Gamma$ and $Z$ is increased by a factor of 4, so that it is larger
than remaining two pockets.  In addition, the $d_{z^2}$ hole pocket is
not present at $Z$ in GGA+A so that now there are only two hole
pockets at the $Z$ point (versus three hole pockets at $Z$ in GGA).
Therefore, a three-dimensional ellipsoidal Fermi surface exists at
$\Gamma$ in GGA+A.

\subsection{Comparison with ARPES}
\label{sec:arpes}

Now we compare modifications in the band structure due to +A term with
the currently available experimental data on
Ba$_{0.6}$K$_{0.4}$Fe$_2$As$_2$ band structure.

First, in angle-resolved photoemission spectroscopy (ARPES)
experiment, three hole pockets are observed at the zone center and the
largest pocket is shown to originate from $d_{xy}$
orbital\cite{evtushinsky} as in our GGA+A calculation.  Second, large
elongation of M point pocket towards the $\Gamma$ and $Z$ points we
find using +A was experimentally observed in angle-resolved
photoemission spectroscopy from Ref.~\onlinecite{malaeb}. Third, the
presence of three-dimensional FS in Ba$_{0.6}$K$_{0.4}$Fe$_2$As$_2$
was suggested from $c$-axis polarized optical
measurements.\cite{cheng} The optical experiment found that the
$c$-axis data only exhibit a small difference across T$_{c}$.  This
indicates the existence of three-dimensional FS with a dispersive band
along the $c$ axis.

\section{Specific heat}
\label{sec:heat}

\begin{figure} 
\epsfig{file=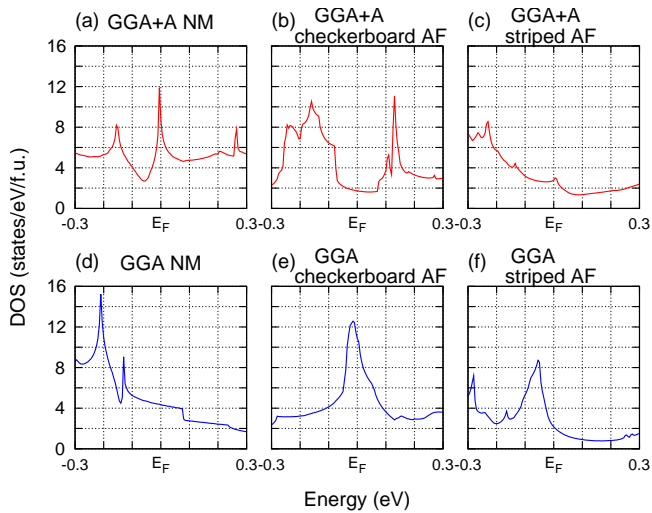,width=8.5cm,angle=0,clip=} 
\caption{
Ba$_{0.6}$K$_{0.4}$Fe$_2$As$_2$ density of states for both spin
components, per two-iron formula unit in the nonmagnetic (NM),
checkerboard antiferromagnetic, and single-stripe
antiferromagnetic state, both in GGA+A (a,b,c) and GGA (d,e,f).
\label{fig:dos}
}
\end{figure}

In this section we discuss the calculated specific heat of
Ba$_{0.6}$K$_{0.4}$Fe$_2$As$_2$.
The specific heat coefficient $\gamma_n$ is defined as,
\begin{equation}
 \gamma_n=(1+\lambda)\gamma_0.
 \label{eq:heat}
\end{equation}
Here $\gamma_0$ is the Sommerfeld coefficient proportional to DOS at
the Fermi energy, and $\lambda$ is the electron-phonon coupling
coefficient.

First we discuss the density of states in GGA and GGA+A.  In GGA the
DOS at the Fermi energy of nonmagnetic Ba$_{0.6}$K$_{0.4}$Fe$_2$As$_2$
is 4.4~states~eV$^{-1}$~f.u.$^{-1}$ (the energy dependence of DOS is
shown in Fig.~\ref{fig:dos}). Similar value (3.1--5.5~states~eV$^{-1}$~f.u.$^{-1}$) for DOS
was found in previous calculations.\cite{singh,shein} 

In our GGA+A calculation, DOS at the peak value near the Fermi level is
11.9~states~eV$^{-1}$~f.u.$^{-1}$,
almost three times larger than in GGA. 
Since $\gamma_0$ is proportional to DOS, it is also increased by a
factor of 3 in GGA+A over GGA (see Table~\ref{tab:heat}).

Increase in the DOS after inclusion of +A term originates from the
changes of the band structure at the $M$ point.  In the GGA+A, the
bottom of the electron-like band at the $M$ point (and the
corresponding van Hove singularity) is placed almost at the $E_{F}$.
Furthermore, the DOS at the van Hove singularity is enhanced due to
the renormalization of the band width and the formation of a saddle
point at the $M$ point [Fig.~\ref{fig:band}(a)].

After having discussed the $\gamma_0$, we now discuss the contribution
of the electron-phonon coupling coefficient ($\lambda$) to the heat
capacity $\gamma_n$.  We calculated the electron-phonon coupling
coefficient $\lambda$ using the Wannier interpolation technique\cite{giustino}
and the Quantum-ESPRESSO package.\cite{giannozzi} 
The electron-phonon coupling in the
nonmagnetic GGA+A calculation is 0.37, about two times larger than
0.18 obtained in GGA (see Table~\ref{tab:heat}). However, the heat
capacity ($\gamma_n$) is proportional to $1+\lambda$ so the increase
in $\lambda$ in GGA+A increases $\gamma_n$ by 16~\%, 
in addition to the dominant increase from larger DOS.

Taking both terms together ($\gamma_0$ and $1+\lambda$) we find that
within GGA+A method specific heat coefficient $\gamma_n$ equals
38~mJ~mol$^{-1}$~K$^{-2}$, which is much closer to the experimentally
measured values (40--50~mJ~mol$^{-1}$~K$^{-2}$) than the GGA
result (12~mJ~mol$^{-1}$~K$^{-2}$).

\begin{table}
\caption{\label{tab:heat} A comparison of the 
Sommerfeld coefficient ($\gamma_0$), electron-phonon coupling
  ($\lambda$), and enhanced normal-state specific heat coefficient
  ($\gamma_n$) of Ba$_{0.6}$K$_{0.4}$Fe$_2$As$_2$ in GGA, GGA+A, and
  from experiment
  (Ref.~\onlinecite{popovich,kant,storey,rotundu}). Coefficients
  $\gamma_0$ and $\gamma_n$ are given in mJ~mol$^{-1}$~K$^{-2}$ in a
  two-iron atom unit cell and for both spin components.}
\begin{ruledtabular} 
\begin{tabular}{lclc} 
 & $\gamma_0$ & $\lambda$ & $\gamma_n$ \\ 
\hline
\multicolumn{4}{l}{Nonmagnetic} \\
\quad \quad GGA+A       &  28 & 0.37             & 38 \\
\quad \quad GGA         &  10 & 0.18\cite{boeri} & 12 \\
\quad \quad Experiment  &     &                  & 40--50 \\
\multicolumn{4}{l}{Checkerboard} \\
\quad \quad GGA+A      & 4.1 & 0.80              & 7.3 \\
\quad \quad GGA        &  26 & 0.33\cite{boeri}  & 34 \\
\multicolumn{4}{l}{Single-stripe} \\
\quad \quad GGA+A      & 6.6 & 0.90              & 12 \\
\quad \quad GGA        & 5.1 & 0.18\cite{boeri}  & 6.0 \\
\end{tabular}
\end{ruledtabular}
\end{table}

\subsection{Antiferromagnetic ground states}

So far we discussed the specific heat in the nonmangetic ground state of
Ba$_{0.6}$K$_{0.4}$Fe$_2$As$_2$, now we consider two antiferromagnetic
ground states: striped and checkerboard. The striped case is especially
important, since this is the experimentally determined ground state of
the parent compound BaFe$_2$As$_2$.  We study the alternative ground state
(checkerboard) for a comparison with the striped phase.

The (single-)stripe order consists of ferromagnetically arranged
chains of iron atoms, with antiferromagnetic aligment between
neighboring chains.  On the other hand, in the checkerboard
antiferromagnetic order magnetic moments on all neighboring iron atoms
in point in opposite directions.

For easier comparison with the nonmagnetic calculations, in our
magnetic GGA+A calculations we use the same value of $A_c$ and the
same function $f(r)$.

In the striped state, the peak in DOS occuring 50~meV below the Fermi
level in GGA is shifted to $E_{F}-$230~meV when +A is included but
there is no significant change in the DOS at the $E_{F}$
[Figs.~\ref{fig:dos}(c) and (f)].  However, in the checkerboard state
within GGA+A we obtain the DOS at $E_{F}$ equal to
1.7~states~eV$^{-1}$~f.u.$^{-1}$, which is about one sixth of the GGA
result (see Figs.~\ref{fig:dos}(b) and (e)).  This suppression in the
checkerboard state is due to the occurrence of a Jahn-Teller
distortion in GGA+A, which is lowering the crystal symmetry from
tetragonal to orthorhombic.

Eventhough within GGA+A DOS at $E_{F}$ is relatively small in the
striped state (2.8~states~eV$^{-1}$~f.u.$^{-1}$) the electron-phonon
coupling is significantly larger than in the nonmagnetic case. We
obtained $\lambda=$~0.90 (see Table~\ref{tab:heat}) in striped state
which is $\sim$60~\% larger than in GGA.  As we said earlier, DOS in
striped state is nearly the same in GGA and GGA+A. Therefore, strong
enhancement of $\lambda$ in GGA+A must originate from other sources,
and not simply from increased DOS. However, the origin of this
enhancement is not the focus of this paper, and it will be reported
elsewhere.

\section{Conclusion}

By increasing the potential on iron atoms, 
making them slightly more repulsive for electrons
significantly improves the structural, magnetic, and electronic properties
of Ba$_{0.6}$K$_{0.4}$Fe$_2$As$_2$, as calculated within DFT.  
The main result of this paper is that with a corrected potential (+A) on iron
atom, the heat capacity of Ba$_{0.6}$K$_{0.4}$Fe$_2$As$_2$ is increased
more than threefold, in good agreement with experimental
data. Applying the same correction to the magnetic states, we find
that electron-phonon coupling is strongly enhanced. This observation
might be crucial in understanding the superconducting properties of
iron-based superconductors.

\section*{Acknowledgements}

We thank Professors N. E. Phillips and R. J. Birgeneau for useful discussions.
This work was supported by National Science Foundation Grant
No. DMR10-1006184 (electronic and magnetic structure calculation) 
and by the Director, Office of Science, Office of
Basic Energy Sciences, Materials Sciences and Engineering Division,
U.S. Department of Energy under Contract No. DE-AC02-05CH11231
(electron-phonon calculation).
Computational resources have been provided by the DOE at Lawrence
Berkeley National Laboratory's NERSC facility.

\scm{I think that if you have two grants we need to say which part of
the work belongs to which grant. One way to divide this is to say that
structure and DOS were done by one grant and that lambda calculation
was done by another grant. I don't know which one should go to which
grant... Marvin?}

\end{document}